\documentclass[cits]{PoS}
\usepackage{amssymb, graphicx, latexsym, amsmath, bbm, color}

\newcommand{\dd}{{\rm{d}}}

\newcommand{\tr}{{\rm Tr}}

\newcommand{\qhat}{\hat{q}}

\newcommand{\gE}{g_{\mbox{\tiny{E}}}}

\newcommand{\mE}{m_{\mbox{\tiny{E}}}}

\newcommand{\eq}{\begin{equation}}
\newcommand{\en}{\end{equation}}

\title{Jet quenching in a strongly interacting plasma -- A lattice approach}

\ShortTitle{Jet quenching in a strongly interacting plasma -- A lattice approach}

\author{\speaker{Marco Panero}\\
        Department of Physics and Helsinki Institute of Physics, University of Helsinki,\\
        P.O. Box 64, FI-00014 Helsinki, Finland\\
        E-mail: \email{marco.panero@helsinki.fi}}

\author{Kari Rummukainen\\
        Department of Physics and Helsinki Institute of Physics, University of Helsinki,\\
        P.O. Box 64, FI-00014 Helsinki, Finland\\
        E-mail: \email{kari.rummukainen@helsinki.fi}}

\author{Andreas Sch\"afer\\
        Institute for Theoretical Physics, University of Regensburg,\\
        D-93040 Regensburg, Germany\\
        E-mail: \email{andreas.schaefer@physik.uni-regensburg.de}}

\abstract{The phenomenon of jet quenching, related to the momentum broadening of a high-energy parton, provides important experimental evidence for the production of a strongly coupled, deconfined medium in heavy-ion collisions. Its theoretical description has been addressed in a number of works, both perturbatively and non-perturbatively (using the gauge-gravity duality). In this contribution, following a proposal by Caron-Huot, we discuss a novel approach to this problem, enabling one to extract non-perturbative information on this real-time phenomenon from simulations on a Euclidean lattice.
\vspace{4cm}
\begin{flushright}
HIP-2013-17/TH
\end{flushright}}

\FullConference{The European Physical Society Conference on High Energy Physics \\
		 18-24 July, 2013\\
		 Stockholm, Sweden}

\begin{document}

\section{Introduction}
More than thirty years ago, Bjorken suggested a possible way to detect the creation of deconfined QCD matter in collisions of ultrarelativistic nuclei: due to interactions with the medium constituents, a hard parton propagating through the quark-gluon plasma (QGP) at a given temperature $T$ would experience energy loss and momentum broadening, and this would result in the suppression of final-state hadrons with large transverse momentum and of back-to-back correlations~\cite{Bjorken:1982qr}. This prediction was eventually confirmed by experiments~\cite{experiments}.

Providing a firm theoretical description for this beautiful physical idea is, however, challenging, as it involves an interplay of both perturbative and non-perturbative physics effects~\cite{CasalderreySolana:2007zz}. Even if one focuses only on the short-distance interactions between the hard parton and QGP constituents~\cite{Baier:1996kr}, the problem is still complicated by the fact that, for temperatures within the reach of present experiments, the QCD coupling $g$ is not very small, so perturbative computations may not be reliable. On the other hand, strong-coupling approaches based on the gauge/string duality, like the ones carried out for a massless hard parton~\cite{Liu:2006ug} or for the drag force experienced by a heavy quark~\cite{heavy_quark_drag_force}, are not based on the QCD Lagrangian. Finally, non-perturbative lattice QCD computations are not straightforward for this real-time problem.

In this contribution, however, we would like to discuss some recent progress in the latter direction~\cite{Panero:2013pla}, based on an idea proposed in ref.~\cite{CaronHuot:2008ni}. Related studies include refs.~\cite{Ghiglieri:2013gia, Laine:2012ht, other_similar_qhat_computations}, whereas a different way to study the jet quenching phenomenon on the lattice was proposed in ref.~\cite{Majumder:2012sh}.

\section{Soft contribution to jet quenching from a Euclidean lattice}
Jet quenching can be described in terms of a phenomenological parameter $\qhat$, defined as the average increase in the squared transverse momentum component $p_\perp$ of the hard parton per unit length. This quantity can be expressed in terms of a differential collisional rate between the parton and plasma constituents $C(p_\perp)$:
\begin{equation}
\qhat = \frac{\langle p^2_\perp \rangle}{L} = \int \frac{ \dd^2 p_\perp}{(2\pi)^2} p^2_\perp C(p_\perp).
\end{equation}
In turn, $C(p_\perp)$ is related to the two-point correlation function of light-cone Wilson lines. Although the full computation of this correlator cannot be carried out on a Euclidean lattice, it is possible to extract the non-perturbative contributions to it from the soft sector, i.e. from physics at momentum scales up to $gT$, which can be proven to be time-independent~\cite{CaronHuot:2008ni, Ghiglieri:2013gia}. Evaluating the non-perturbative contribution from soft (and ultrasoft, of order $g^2 T/\pi$) modes is important, since they are responsible for the peculiar analytical structure of weak-coupling computations in thermal QCD and for the large corrections affecting the corresponding perturbative series. A proper systematic framework to deal with these problems can be formulated in terms of dimensionally reduced effective theories~\cite{dim_red}. In particular, the soft-scale dynamics can be described by electrostatic QCD (EQCD): an effective theory for the static QGP modes, given by three-dimensional Yang-Mills theory coupled to an adjoint scalar field,
\begin{equation}
\mathcal{L} = \frac{1}{4} F_{ij}^a F_{ij}^a + \tr \left( (D_i A_0)^2 \right) + \mE^2 \tr \left( A_0^2 \right) + \lambda_3 \left( \tr \left( A_0^2 \right) \right)^2.
\end{equation}
Its parameters (the 3D gauge coupling $\gE$ and the mass- and quartic-term coefficients) can be fixed by \emph{matching} to the physics of high-temperature QCD, and the theory can be regularized on a lattice. We chose a setup corresponding to the dimensional reduction of QCD with $n_f=2$ light dynamical quark flavors, at two temperatures ($T \simeq 398$~MeV and $2$~GeV) approximately twice and ten times larger than the deconfinement temperature~\cite{Hietanen:2008tv}.

\begin{figure}[-t]
\centerline{\includegraphics[height=0.25\textheight]{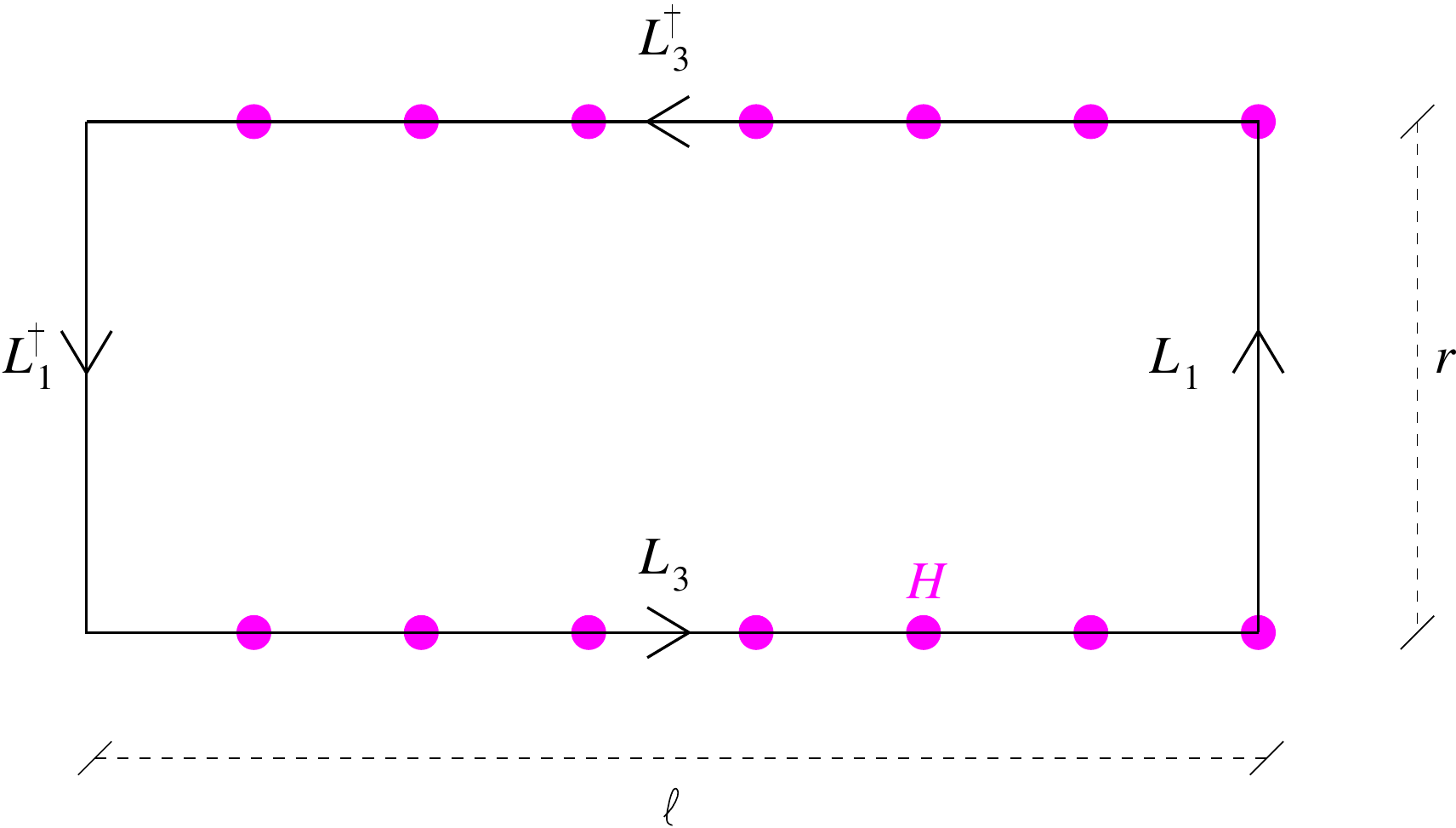}}
\caption{The ``decorated'' Wilson loop $W(\ell, r)$ describing a two-point correlation function of light-cone Wilson lines involves Hermitian parallel transporters $H(x)$ along the real-time direction.\label{fig:operator}}
\end{figure}

We simulate this theory and study (a gauge-invariant generalization of) the two-point correlator of light-cone Wilson lines, defined in terms of a lattice operator which involves parallel transporters $H(x) = \exp [- a \gE^2 A_0(x) ]$ along \emph{real time}, which are Hermitian---rather than unitary---operators. This results in a ``decorated'' Wilson loop $W(\ell, r)$ (see fig.~\ref{fig:operator}) with well-defined renormalization properties~\cite{D'Onofrio_et_al}. From its expectation values (computed with the multilevel algorithm~\cite{Luscher:2001up}) we extract a ``potential''
\begin{equation}
V(r) = -\frac{1}{\ell} \ln\langle W(\ell, r) \rangle,
\end{equation}
which is equal to the transverse Fourier transform of $C(p_\perp)$.

\begin{figure}[-t]
\centerline{\includegraphics[height=0.25\textheight]{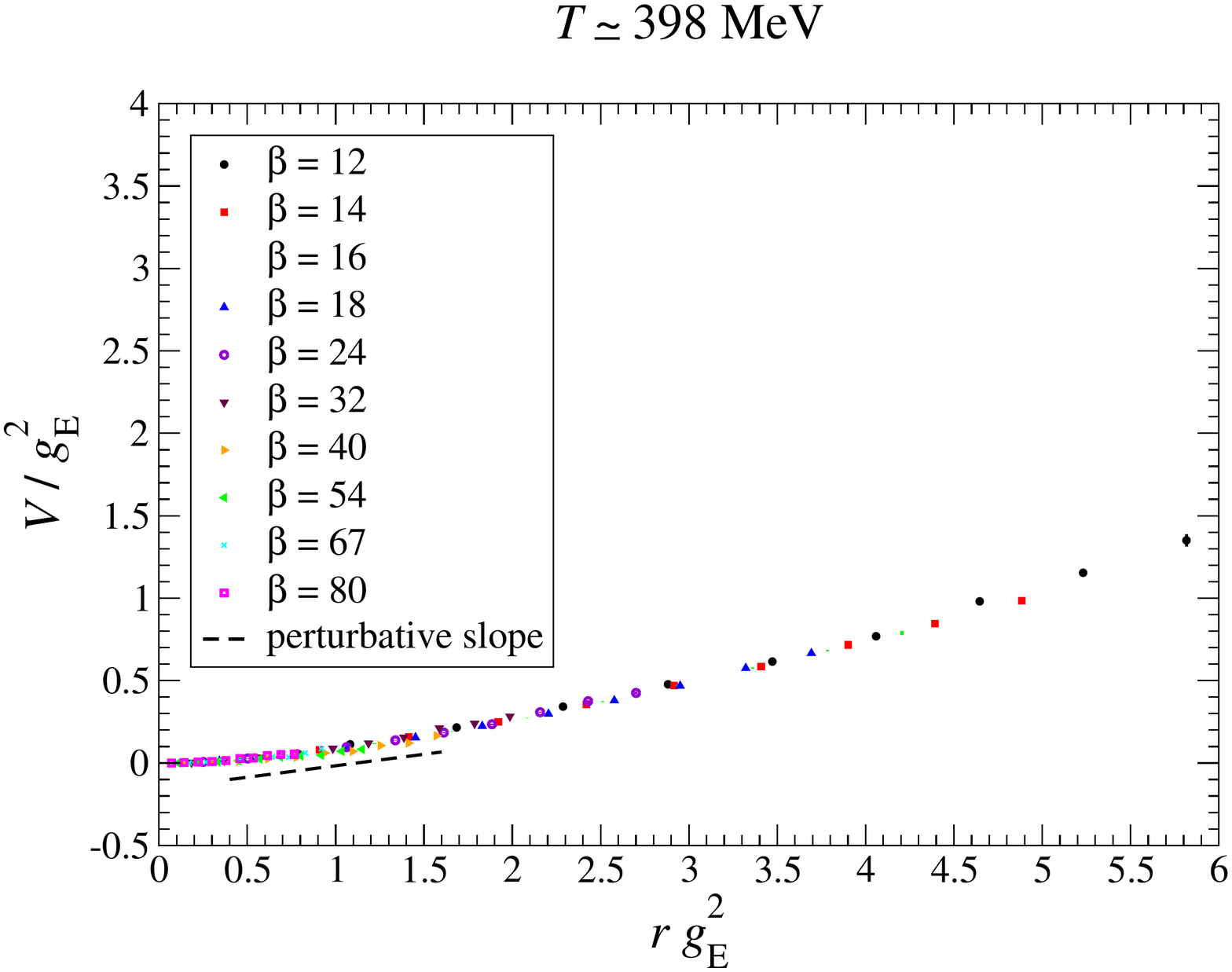} \hfill \includegraphics[height=0.25\textheight]{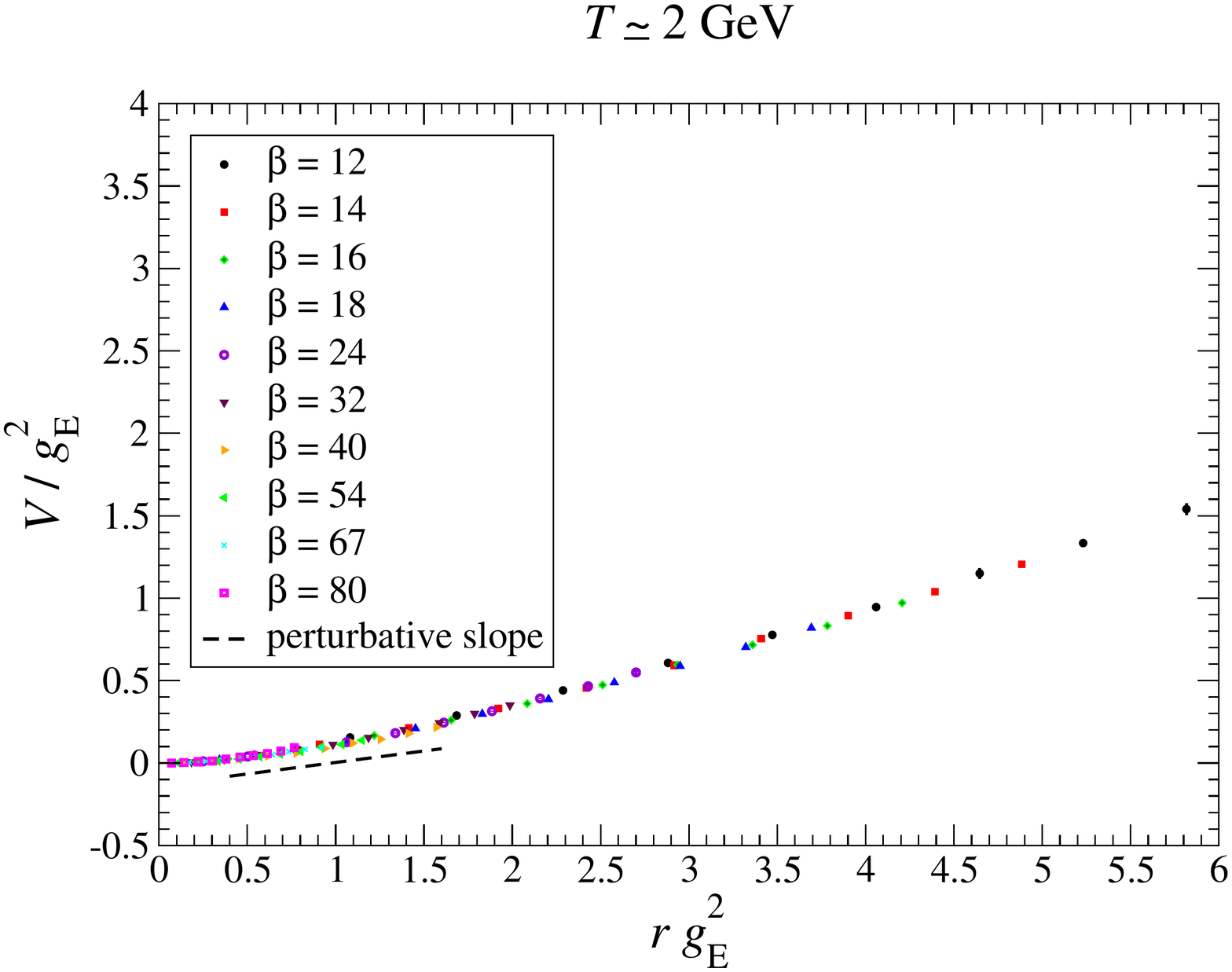}}
\caption{The ``potential'' $V(r)$ extracted from the expectation values of $W(\ell, r)$, at $T \simeq 398$~MeV (left-hand-side panel) and at $T \simeq 2$~GeV (right-hand-side panel). Both $V$ and $r$ are shown in the appropriate units of the dimensionful 3D gauge coupling $\gE$. The slope predicted perturbatively for the potential at values $r \gE^2 \gtrsim 1$ is also displayed.
\label{fig:V}}
\end{figure}

At short distances our results for $V(r)$ (shown in fig.~\ref{fig:V}) are compatible with perturbative expectations, which involve, in particular, a delicate cancellation between gluon and scalar propagators~\cite{CaronHuot:2008ni, Ghiglieri:2013gia}. The non-perturbative contributions to $V(r)$ can be related to $\qhat$: the latter is given by the second moment of the distribution associated with $C(p_\perp)$, which corresponds to curvature terms in $V(r)$. Following an approach similar to ref.~\cite{Laine:2012ht}, we arrive at quite large values for the soft NLO contribution to the jet quenching parameter: $0.55(5)\gE^6$ for $T \simeq 398$~MeV, and $0.45(5)\gE^6$ for $T \simeq 2$~GeV. In turn, these numbers lead to a final estimate for $\qhat$ around $6$~GeV$^2$/fm for RHIC temperatures, comparable with those from holographic estimates~\cite{Liu:2006ug} and from computations with phenomenological input~\cite{Dainese_Eskola}.

\section{Conclusions and outlook}
We have shown that, contrary to na\"{\i}ve intuition, the lattice study of certain real-time phenomena involving physics on the light cone is possible. Here we have discussed the phenomenon of jet quenching in thermal QCD, but related ideas have also been proposed for QCD at zero temperature~\cite{Ji_Lin}.

By construction, the bosonic effective theory that we simulated in our approach allows one to separate the soft contributions to $\qhat$ from those due to hard thermal modes, with momenta of order $\pi T$. It does so in a controlled, systematic way, consistent with the modern theoretical framework to study finite-temperature QCD~\cite{dim_red, Laine:2005ai}.

In the near future, we plan to improve our extrapolation of the potential $V(r)$ to the continuum limit at short $r$ by carrying out further simulations on finer lattices, and/or using improved actions~\cite{Mykkanen:2012dv}. It would also be interesting to study the dependence of $\qhat$ on $T$, and on the number of color charges $N$. As it is well-known, the large-$N$ limit is characterized by a rich and interesting phenomenology~\cite{large_N_general}, and lattice studies have shown that the static equilibrium properties of the QGP have very little dependence on $N$, both in four~\cite{large_N_finite_T_4D} and in three~\cite{large_N_finite_T_3D} spacetime dimensions.

\vskip1.0cm 
\noindent{\bf Acknowledgements}\\
This work is supported by the Academy of Finland (project 1134018) and by the German DFG (SFB/TR 55). M.P. acknowledges financial support for travel from Universit\`a della Calabria (fund PAPA~340101 -- Contributo Unical~2012 Progetto FIRB), Cosenza, Italy. Part of the numerical simulations was carried out at the Finnish IT Center for Science (CSC) in Espoo, Finland.


\begin{thebibliography}{99}

\bibitem{Bjorken:1982qr}
  J.~D.~Bjorken,
  \emph{Highly Relativistic Nucleus-Nucleus Collisions: The Central Rapidity Region},
  \emph{Phys.\ Rev.\ D} {\bf 27} (1983) 140.

\bibitem{experiments}
  K.~Adcox {\it et al.}  [PHENIX Collaboration],
  \emph{Suppression of hadrons with large transverse momentum in central Au+Au collisions at $\sqrt{s_{NN}}$ = 130~GeV},
  \emph{Phys.\ Rev.\ Lett.}\  {\bf 88} (2002) 022301
  {\tt [nucl-ex/0109003]}.
%
  I.~Arsene {\it et al.}  [BRAHMS Collaboration],
  \emph{Transverse momentum spectra in Au+Au and d+Au collisions at $\sqrt{s_{NN}}$ = 200~GeV and the pseudorapidity dependence of high p(T) suppression},
  \emph{Phys.\ Rev.\ Lett.}\  {\bf 91} (2003) 072305
  {\tt [nucl-ex/0307003]}.
%
  G.~Aad {\it et al.}  [ATLAS Collaboration],
  \emph{Observation of a Centrality-Dependent Dijet Asymmetry in Lead-Lead Collisions at $\sqrt{s_{NN}}=2.77$~TeV with the ATLAS Detector at the LHC},
  \emph{Phys.\ Rev.\ Lett.}\  {\bf 105} (2010) 252303
  {\tt [arXiv:1011.6182 [hep-ex]]}.
%
  K.~Aamodt {\it et al.}  [ALICE Collaboration],
  \emph{Suppression of Charged Particle Production at Large Transverse Momentum in Central Pb--Pb Collisions at $\sqrt{s_{NN}} = 2.76$~TeV},
  \emph{Phys.\ Lett.\ B} {\bf 696} (2011) 30
  {\tt [arXiv:1012.1004 [nucl-ex]]}.
%
  S.~Chatrchyan {\it et al.}  [CMS Collaboration],
  \emph{Observation and studies of jet quenching in PbPb collisions at nucleon-nucleon center-of-mass energy = 2.76~TeV},
  \emph{Phys.\ Rev.\ C} {\bf 84} (2011) 024906
  {\tt [arXiv:1102.1957 [nucl-ex]]}.

\bibitem{CasalderreySolana:2007zz}
  J.~Casalderrey-Solana and C.~A.~Salgado,
  \emph{Introductory lectures on jet quenching in heavy ion collisions,}
  \emph{Acta Phys.\ Polon.\ B} {\bf 38} (2007) 3731
  {\tt [arXiv:0712.3443 [hep-ph]]}.

\bibitem{Baier:1996kr}
  R.~Baier \emph{et al.}, 
  \emph{Radiative energy loss of high-energy quarks and gluons in a finite volume quark - gluon plasma},
  \emph{Nucl.\ Phys.\ B} {\bf 483} (1997) 291
  {\tt [hep-ph/9607355]}.

\bibitem{Liu:2006ug}
  H.~Liu, K.~Rajagopal and U.~A.~Wiedemann,
  \emph{Calculating the jet quenching parameter from AdS/CFT},
  \emph{Phys.\ Rev.\ Lett.}\  {\bf 97} (2006) 182301
  {\tt [hep-ph/0605178]}.

\bibitem{heavy_quark_drag_force}
  C.~P.~Herzog \emph{et al.}, 
  \emph{Energy loss of a heavy quark moving through N=4 supersymmetric Yang-Mills plasma},
  \emph{JHEP} {\bf 0607} (2006) 013
  {\tt [hep-th/0605158]}.
%
  S.~S.~Gubser,
  \emph{Drag force in AdS/CFT},
  \emph{Phys.\ Rev.\ D} {\bf 74} (2006) 126005
  {\tt [hep-th/0605182]}.
%
  J.~Casalderrey-Solana and D.~Teaney,
  \emph{Heavy quark diffusion in strongly coupled N=4 Yang-Mills},
  \emph{Phys.\ Rev.\ D} {\bf 74} (2006) 085012
  {\tt [hep-ph/0605199]}.
%
  U.~G{\"u}rsoy \emph{et al.}, 
  \emph{Thermal Transport and Drag Force in Improved Holographic QCD},
  \emph{JHEP} {\bf 0912} (2009) 056
  {\tt [arXiv:0906.1890 [hep-ph]]}.

\bibitem{Panero:2013pla}
  M.~Panero, K.~Rummukainen and A.~Sch\"afer,
  \emph{A lattice study of the jet quenching parameter},
  {\tt arXiv:1307.5850 [hep-ph]}.
  See also
  M.~Panero, K.~Rummukainen and A.~Sch\"afer,
  \emph{Momentum broadening of partons on the light cone from the lattice},
  \pos{PoS(LATTICE 2013)173}
  {\tt [arXiv:1309.3212 [hep-lat]]}
  for a concise summary.

\bibitem{CaronHuot:2008ni}
  S.~Caron-Huot,
  \emph{O(g) plasma effects in jet quenching},
  \emph{Phys.\ Rev.\ D} {\bf 79} (2009) 065039
  {\tt [arXiv:0811.1603 [hep-ph]]}.

\bibitem{Ghiglieri:2013gia}
  J.~Ghiglieri \emph{et al.}, 
  \emph{Next-to-leading order thermal photon production in a weakly coupled quark-gluon plasma},
  \emph{JHEP} {\bf 1305} (2013) 010
  {\tt [arXiv:1302.5970 [hep-ph]]}.

\bibitem{Laine:2012ht}
  M.~Laine,
  \emph{A non-perturbative contribution to jet quenching},
  \emph{Eur.\ Phys.\ J.\ C} {\bf 72} (2012) 2233
  {\tt [arXiv:1208.5707 [hep-ph]]}.

\bibitem{other_similar_qhat_computations}
  M.~Benzke \emph{et al.}, 
  \emph{Gauge invariant definition of the jet quenching parameter},
  \emph{JHEP} {\bf 1302} (2013) 129
  {\tt [arXiv:1208.4253 [hep-ph]]}.
%
  M.~Laine and A.~Rothkopf,
  \emph{Light-cone Wilson loop in classical lattice gauge theory},
  \emph{JHEP} {\bf 1307} (2013) 082
  {\tt [arXiv:1304.4443 [hep-ph]]};
%
  \emph{Towards understanding thermal jet quenching via lattice simulations},
  \pos{PoS(LATTICE 2013)174}.
%
  I.~O.~Cherednikov, J.~Lauwers and P.~Taels,
  \emph{On a Wilson lines approach to the study of jet quenching},
  {\tt [arXiv:1307.5518 [hep-ph]]}.

\bibitem{Majumder:2012sh}
  A.~Majumder,
  \emph{Calculating the Jet Quenching Parameter $\hat{q}$ in Lattice Gauge Theory},
  \emph{Phys.\ Rev.\ C} {\bf 87} (2013) 034905
  {\tt [arXiv:1202.5295 [nucl-th]]}.

\bibitem{dim_red}
  E.~Braaten and A.~Nieto,
  \emph{Effective field theory approach to high temperature thermodynamics},
  \emph{Phys.\ Rev.\ D} {\bf 51} (1995) 6990
  {\tt [hep-ph/9501375]};
%
  \emph{Free energy of QCD at high temperature},
  \emph{Phys.\ Rev.\ D} {\bf 53} (1996) 3421
  {\tt [hep-ph/9510408]}.
%
  K.~Kajantie \emph{et al.}, 
  \emph{Generic rules for high temperature dimensional reduction and their application to the standard model},
  \emph{Nucl.\ Phys.\ B} {\bf 458} (1996) 90
  {\tt [hep-ph/9508379]}.

\bibitem{Hietanen:2008tv}
  A.~Hietanen \emph{et al.}, 
  \emph{Three-dimensional physics and the pressure of hot QCD},
  \emph{Phys.\ Rev.\ D} {\bf 79} (2009) 045018
  {\tt [arXiv:0811.4664 [hep-lat]]}.

\bibitem{D'Onofrio_et_al}
  M.~D'Onofrio \emph{et al.},
  to appear.

\bibitem{Luscher:2001up}
  M.~L\"uscher and P.~Weisz,
  \emph{Locality and exponential error reduction in numerical lattice gauge theory},
  \emph{JHEP} {\bf 0109} (2001) 010
  {\tt [hep-lat/0108014]}.

\bibitem{Dainese_Eskola}
  K.~J.~Eskola \emph{et al.}, 
  \emph{The Fragility of high-p(T) hadron spectra as a hard probe},
  \emph{Nucl.\ Phys.\ A} {\bf 747} (2005) 511
  {\tt [hep-ph/0406319]}.
%
  A.~Dainese, C.~Loizides and G.~Pai\'c,
  \emph{Leading-particle suppression in high energy nucleus-nucleus collisions},
  \emph{Eur.\ Phys.\ J.\ C} {\bf 38} (2005) 461
  {\tt [hep-ph/0406201]}.

\bibitem{Ji_Lin}
  X.~Ji,
  \emph{Parton Physics on Euclidean Lattice},
  {\tt arXiv:1305.1539 [hep-ph]}.
%
  H.~W.~Lin,
  \emph{Calculating the $x$ Dependence of Nucleon Parton Distribution Functions},
  \pos{PoS(LATTICE 2013)293}.

\bibitem{Laine:2005ai}
  M.~Laine and Y.~Schr\"oder,
  \emph{Two-loop QCD gauge coupling at high temperatures},
  \emph{JHEP} {\bf 0503} (2005) 067
  {\tt [hep-ph/0503061]}.

\bibitem{Mykkanen:2012dv}
  A.~Mykk\"anen,
  \emph{The static quark potential from a multilevel algorithm for the improved gauge action},
  \emph{JHEP} {\bf 1212} (2012) 069
  {\tt [arXiv:1209.2372 [hep-lat]]}.

\bibitem{large_N_general}
  G.~'t Hooft,
  \emph{A Planar Diagram Theory for Strong Interactions},
  \emph{Nucl.\ Phys.\ B} {\bf 72} (1974) 461.
  See also
%
  B.~Lucini and M.~Panero,
  \emph{SU(N) gauge theories at large N},
  \emph{Phys.\ Rept.}\  {\bf 526} (2013) 93
  {\tt [arXiv:1210.4997 [hep-th]]};
  \emph{Introductory lectures to large-N QCD phenomenology and lattice results},
  {\tt arXiv:1309.3638 [hep-th]]}
  and 
%
  M.~Panero,
  \emph{Recent results in large-N lattice gauge theories},
  \pos{PoS(LATTICE 2012)010}
  {\tt [arXiv:1210.5510 [hep-lat]]}
  for recent reviews.

\bibitem{large_N_finite_T_4D}
  B.~Lucini, M.~Teper and U.~Wenger,
  \emph{Properties of the deconfining phase transition in SU(N) gauge theories},
  \emph{JHEP} {\bf 0502} (2005) 033
  {\tt [hep-lat/0502003]}.
%
  B.~Bringoltz and M.~Teper,
  \emph{The Pressure of the SU(N) lattice gauge theory at large-N},
  \emph{Phys.\ Lett.\ B} {\bf 628} (2005) 113
  {\tt [hep-lat/0506034]}.
%
  S.~Datta and S.~Gupta,
  \emph{Scaling and the continuum limit of the finite temperature deconfinement transition in SU$(N_c)$ pure gauge theory},
  \emph{Phys.\ Rev.\ D} {\bf 80} (2009) 114504
  {\tt [arXiv:0909.5591 [hep-lat]]}.
%
  M.~Panero,
  \emph{Thermodynamics of the QCD plasma and the large-N limit},
  \emph{Phys.\ Rev.\ Lett.}\  {\bf 103} (2009) 232001
  {\tt [arXiv:0907.3719 [hep-lat]]}.
%
  A.~Mykk\"anen, M.~Panero and K.~Rummukainen,
  \emph{Casimir scaling and renormalization of Polyakov loops in large-N gauge theories},
  \emph{JHEP} {\bf 1205} (2012) 069
  {\tt [arXiv:1202.2762 [hep-lat]]}.
%
  B.~Lucini, A.~Rago and E.~Rinaldi,
  \emph{SU($N_c$) gauge theories at deconfinement},
  \emph{Phys.\ Lett.\ B} {\bf 712} (2012) 279
  {\tt [arXiv:1202.6684 [hep-lat]]}.
 
\bibitem{large_N_finite_T_3D}
  M.~Caselle \emph{et al.}, 
  \emph{Thermodynamics of SU(N) Yang-Mills theories in 2+1 dimensions I -- The confining phase},
  \emph{JHEP} {\bf 1106} (2011) 142
  {\tt [arXiv:1105.0359 [hep-lat]]};
%
  \emph{Thermodynamics of SU(N) Yang-Mills theories in 2+1 dimensions II -- The deconfined phase},
  \emph{JHEP} {\bf 1205} (2012) 135
  {\tt [arXiv:1111.0580 [hep-th]]}.
%
  P.~Bialas \emph{et al.}, 
  \emph{Three dimensional finite temperature SU(3) gauge theory near the phase transition},
  \emph{Nucl.\ Phys.\ B} {\bf 871} (2013) 111
  {\tt [arXiv:1211.3304 [hep-lat]]}.

\end{thebibliography}
\end{document}